\documentclass{sigchi}

\toappear{\scriptsize Permission to make digital or hard copies of all or part of this work for personal or classroom use is granted without fee provided that copies are not made or distributed for profit or commercial advantage and that copies bear this notice and the full citation on the first page. Copyrights for components of this work owned by others than the author(s) must be honored. Abstracting with credit is permitted. To copy otherwise, or republish, to post on servers or to redistribute to lists, requires prior specific permission and/or a fee. Request permissions from permissions@acm.org. \\
{\emph{DIS '20, July 6--10, 2020, Eindhoven, Netherlands.} } \\
\copyright~2020 Copyright is held by the owner/author(s). Publication rights licensed to ACM. \\
ACM ISBN 978-1-4503-6974-9/20/07\ ...\$15.00.\\
http://dx.doi.org/10.1145/3357236.3395501}

\usepackage{balance}
\usepackage{graphics}
\usepackage[T1]{fontenc}
\usepackage{txfonts}
\usepackage{mathptmx}
\usepackage{color}
\usepackage{booktabs}
\usepackage{tabularx}
\usepackage{gensymb}
\usepackage{textcomp}
\usepackage{microtype}
\usepackage{enumitem}
\usepackage[pdflang={en-US},pdftex]{hyperref}

\def\plaintitle{Strangers in the Room: Unpacking Perceptions of `Smartness' and Related Ethical Concerns in the Home}
\def\plainauthor{William Seymour, Reuben Binns, Petr Slovak, Max Van Kleek, Nigel Shadbolt}

\def\plainkeywords{Smart Homes, Smart Devices, Mental Models, Folk Theories}

\makeatletter
\def\url@leostyle{%
  \@ifundefined{selectfont}{
    \def\UrlFont{\sf}
  }{
    \def\UrlFont{\small\bf\ttfamily}
  }}
\makeatother
\def\pprw{8.5in}
\def\pprh{11in}

\definecolor{linkColor}{RGB}{6,125,233}
\hypersetup{%
  pdftitle={\plaintitle},
  pdfauthor={\plainauthor},
  pdfkeywords={\plainkeywords},
  pdfdisplaydoctitle=true,
  bookmarksnumbered,
  pdfstartview={FitH},
  colorlinks,
  citecolor=black,
  filecolor=black,
  linkcolor=black,
  urlcolor=linkColor,
  breaklinks=true,
  hypertexnames=false
}

\begin{document}

\title{\plaintitle}

\numberofauthors{5}
\author{%
  \alignauthor{William Seymour\\
    \affaddr{University of Oxford}\\
    \affaddr{Oxford, UK}\\
    \email{william.seymour@cs.ox.ac.uk}}\\
  \alignauthor{Reuben Binns\\
    \affaddr{University of Oxford}\\
    \affaddr{Oxford, UK}\\
    \email{reuben.binns@cs.ox.ac.uk}}\\
  \alignauthor{Petr Slovak\\
    \affaddr{King's College London}\\
    \affaddr{London, UK}\\
    \email{petr.slovak@kcl.ac.uk}}\\
  \alignauthor{Max Van Kleek\\
    \affaddr{University of Oxford}\\
    \affaddr{Oxford, UK}\\
    \email{max.van.kleek@cs.ox.ac.uk}}\\
 \alignauthor{Nigel Shadbolt\\
    \affaddr{University of Oxford}\\
    \affaddr{Oxford, UK}\\
    \email{nigel.shadbolt@cs.ox.ac.uk}}\\
}

\maketitle

\begin{abstract}
The increasingly widespread use of `smart' devices has raised multifarious ethical concerns regarding their use in domestic spaces. Previous work examining such ethical dimensions has typically either involved empirical studies of concerns raised by specific devices and use contexts, or alternatively expounded on abstract concepts like autonomy, privacy or trust in relation to `smart homes' in general. This paper attempts to bridge these approaches by asking what features of smart devices users consider as rendering them `smart' and how these relate to ethical concerns. Through a multimethod investigation including surveys with smart device users (n=120) and semi-structured interviews (n=15), we identify and describe eight types of smartness and explore how they engender a variety of ethical concerns including privacy, autonomy, and disruption of the social order. We argue that this middle ground, between concerns arising from particular devices and more abstract ethical concepts, can better anticipate potential ethical concerns regarding smart devices.
\end{abstract}





\maketitle

\section{Introduction}
After decades of existence in popular imagination, the notion of a `smart home' as a domestic space enriched with connected digital devices is gradually becoming a reality. Devices such as thermostats that learn your preferred temperature, and speakers you can interact with through conversation, are designed to wrap up `smartness' in a familiar dumb packaging.

Given that many people are at least generally aware of one or more smart home devices, one might think it would be obvious what qualifies them to bear the smart label. While early research into smart devices tended to focus on the broad potential for smart devices to ``imbue homes with intelligence''~\cite{taylor2007homes}, many more recent contributions partition the space into overlapping but distinct definitions of \textit{autonomous systems}~\cite{nilsson2019breaching}, \textit{home assistants}~\cite{fruchter2018consumer}, or more narrow definitions describing single types of device; definitions of smart home devices that go beyond `internet connected'~\cite{zeng2017end} or providing `smart home service functions'~\cite{kim2019definitions} are few and far between.

At the same time, there are very real ethical concerns about the privacy, autonomy, transparency, and social effects of these devices once they are integrated into everyday life. The home as an environment contains finely balanced social equilibria that technology is apt to disrupt. Power dynamics within familial units and between cohabitants are altered by technology that acts as a gatekeeper of information, and for many, the home is the most private space in their lives, a place where they do not expect to be surveilled.

These concerns are often cited at a low level in relation to specific products or deployment settings, or at a high level, considering the ethical implications of smart homes in general. Neither approach adequately considers the broader features shared by multiple devices from a variety of categories and contexts, and while useful for uncovering the implications of smart device usage \emph{in situ}, or ethical concerns writ large, may be of less use in understanding how different constituent elements of smartness, such as automation, or remote control, operate across different contexts to generate ethical concerns.

Instead, this work aims to explore an intermediate level of analysis, making the following contributions: (1) it presents user formulations of smartness: how non-experts consider smartness in the context of smart home devices, and how these concepts of smartness interact across categories of devices; (2) it analyses how these formulations of smartness relate to ethical concerns expressed in the academic literature, as well as how ethical concerns can be caused or ameliorated by features of smart devices and formulations of smartness; (3) it demonstrates how these unpacked formulations of smartness can both reveal and bridge ethical concerns between specific contexts, devices, and locations, in a way not possible through highly specific or very broad approaches.

This is important because these formulations of smartness can help establish a design space of capabilities and user expectations, usable by creators of future smart home systems to guide the design of new `smart' things. Identifying how potential ethical concerns align with specific notions of smartness can also help designers more easily identify and mitigate such concerns, especially where these concerns transcend particular device/context pairings.

\section{Background}
While visions of the smart home have been driven primarily by technologists, it has also been a significant phenomenon of study for social scientists, HCI researchers, and others, for whom the pertinent questions include: what kind of people have or want smart homes, for what purposes, and what are their hopes, concerns, and needs? In a review of academic literature on smart homes, Wilson et al. argue that smart home research tends to take a view of the smart home as either: `functional', i.e. by contribution to better living; `instrumental' towards some societal goal like reduction of energy consumption to help the environment; and a `socio-technical' view, which sees the smart home as a site of social and technical co-evolution, where the use and meaning of new technologies becomes socially constructed~\cite{wilson2015smart}.

A significant portion of such work is (explicitly or implicitly) motivated by questions of adoption; i.e., identifying factors which might lead to people using or rejecting smart devices in their home. Within this framing, ethical concerns are often lumped in with a range of other factors seen as  `barriers to adoption' (a phrase often used throughout the literature, e.g.~\cite{brush2011home}). As Jensen et al argue, (drawing from~\cite{nelson2000case}), \emph{ethics} can be defined as that which relates to how an object \emph{ought} to be designed according to `ethical or moral codes', and is one of three broad kinds of desiderata for smart homes, alongside \emph{reason} (relating to a smart home device's `purpose'), and \emph{aesthetics} (relating to how pleasurable it is to use)~\cite{jensen2018designing}.

\subsection{Smartness as a Historical, Marketing, and Ideological Concept}
As with any loosely defined technological concept (e.g. cloud computing or blockchain), narratives surrounding smartness and what it means for devices to be smart differ greatly based on context and culture. Smartphones, for example, have seen tropes of integration and dis-integration, pitching the ideal of the connected and productive phone user against fears of becoming addicted or out of touch~\cite{harmon2013stories}. The public/private sector balance that drives smart cities, on the other hand, often tends towards more hegemonic discourse, ignoring risks such as technological lock-in and the overreach of surveillance and profiling~\cite{kitchin2015making}.

% unlike the dreams, current smart devices aren't there yet (TODO maybe)
Similar to the smartphone, smart home devices are largely produced by the private sector, although perception differs in that smart home devices are mainly seen as novelties or conveniences, rather than enabling a new way of living to which the smart home is integral. Similarly to with smart cities, the neo-liberal approach to smart device functionality continually situates devices as solutions, rather than as problems, a process which Kitchin describes as designed to ``bring [detractors] into the fold while keeping [vendors'] central mission of capital accumulation and technocratic governance intact''~\cite{kitchin2015making}. But this approach of addressing symptoms rather than the cause of problems only suffices while the pace of development prevents evaluation of past attempts; it is unlikely that introducing a `please and thank you' mode for Alexa, for example, addresses the fundamental concerns around raising children in an environment with voice assistants, but it will likely be obsolete by the time this is fully explored.

Central to all of these narratives (smart phones, cities, and homes) is the echoing of the hopes and fears of contemporary society, often to the detriment of more nuanced accounts of how people use technology. The stories told by designers and marketers tend to produce categories that pitch practices and values against each other (e.g. Harmon's tropes of integration and dis-integration), creating false dichotomies for users of those technologies~\cite{harmon2013stories}, and designers themselves become caught up in the smartness dream, describing attributes that do not exist in contemporary devices (such as contextual awareness and wide-ranging interoperability) when asked what it means to be smart~\cite{kim2019definitions}. How these tropes are perceived matters beyond the realms of marketing and adoption; smart devices increasingly blur previously well-defined categories such as ``human'' and ``machine'', leaving them constantly in flux~\cite{leahu2013categories}. This drives users to `purify' their definitions of these categories in order to keep them disjoint (see also~\cite{reeves1996media}). In this way, we can think of smart homes as approaching ``humanness and machine-ness as intra-actively produced''---Leahu notes that like users, researchers also exhibit this behaviour, naturally over-focussing on certain phenomena when looking for categories~\cite{leahu2013categories}.

\subsection{Empirical Investigation of Situated Smart Home Ethical Concerns}
A significant body of empirical work has investigated ethical concerns, typically through the use of quantitative surveys and experiments, qualitative interviews, and mixed-methods studies, to explore the attitudes and reactions of current (or potential) smart home users. Many such studies are based on small-scale, longitudinal deployments of particular smart devices or lab-based studies. Oulasvirta et al.~\cite{oulasvirta2012long} and Choe et al.~\cite{choe2011living}, for example, used deployed sensors to capture user perceptions over an extended period of time, an approach which was effective at exploring the cognitive effects of living with continuous sensing.

Desjardins et al. explore alternative ways of designing IoT systems for non-stereotypical homes, examining how IoT devices can interact with porous boundaries, temporality, neighbourly relations, agency, and imagined uses for the home environment~\cite{desjardins2019alternative}. Others studies consider households in which devices have already been embedded; for instance, in a 2007 study which pre-dates the more recent wave of smart devices, Woodruff et al. examined a range of simple home automation tools used by orthodox Jewish households to enable the Sabbath day ritual of abstinence from direct interaction with technology~\cite{woodruff2007sabbath}, and other contributions have addressed ethical concerns raised by data collection in the home without a specific focus on smart devices~\cite{goulden2017living, seymour2020informing}.

More recently released classes of smart devices are also of concern, including smart toys~\cite{mcreynolds2017toys}, smart TVs~\cite{malkin2018can}, robotic home assistants~\cite{urquhart2019responsible}, and the range of issues they raise. Studies addressing the use of smart energy grid technology~\cite{goulden2014smart}, and smart meter agents~\cite{costanza2014doing,rodden2013home}, have uncovered a range of concerns ranging from the potential loss of autonomy in the ordering of domestic life resulting from the possibility of delegation, to the impact on (often gendered) power dynamics within households~\cite{matthews2017stories}.

Similarly, in a home deployment of voice-controlled smart speakers, Porcheron et al. found that members of the household transferred their existing levels of authority to interactions with devices~\cite{porcheron2018voice}. When voice-controls are manifested as personified virtual assistants, they may themselves become perceived as social actors, with end-users characterising them with character traits such as `my best friend', `caring', or `commanding'~\cite{mennicken2016s,purington2017alexa}. Voice control can also be an enabling technology, assisting access to information and making everyday tasks easier (such as for the visually impaired~\cite{pradhan2018accessibility} or children on the autistic spectrum~\cite{frauenberger2019thinking}).

\subsection{Expository Work on Smart Home Ethical Principles}
% Fritsch et al. analyse ethical design manifestos for IoT~\cite{fritsch2018calling}, showing how they rally designers and developers around a common cause at the expense of calling for political or judicial change.
In addition to empirical research which elicits concerns pertaining to particular devices in particular contexts (whether real or speculative), a range of expository work addresses smart home ethics from the level of policy, principles, philosophy, cultural critique, and other perspectives. For instance, discussions about the application of legal regulations like data protection~\cite{wachter2018normative}, or `ethical design manifestos' (such as those analysed in~\cite{fritsch2018calling}) tend to consider ethical issues independently from a `bird's eye view', rather than working ground-up from empirical investigation of user concerns. Within such work, overarching principles and forms of harm are enumerated, such as non-discrimination, privacy and accountability (e.g.~\cite{berman2017social, mittelstadt2017designing}).

Such work is useful in situating the ethics of smart homes in relation to more esoteric and theoretical discourse around long-standing moral principles and human rights, as well as drawing points of connection to potentially legally enforceable rights and duties (such as in the Databox project~\cite{crabtree2018building}). As legal obligations are increasingly mandated to be embedded in technologies from the outset, through requirements like privacy- and data protection-by-design~\cite{urquhart2017new}, such work can potentially guide which principles might apply in a range of smart home contexts. However, they may not always reflect the concerns that are in fact elicited or expressed by end-users in reality, and they don't by themselves map how the design space of smart devices might relate to those ethical principles.

\subsection{Between Situated Ethical Concerns and Abstract Ethical Principles}
%In recognition of the widespread ethical concerns about smart devices, a range of tools have been created to help designers.  Cards sets like KnowCards\footnote{\url{github.com/betteriot/betteriot-knowcards}} and Envisioning Cards\footnote{\url{www.envisioningcards.com}} can help to bridge these abstract concerns with concrete scenarios, helping designers to engage with the people they are designing for. 

%Certifications like the Trustable Technology Mark\footnote{\url{https://trustabletech.org}} also help to communicate these processes to consumers, differentiating devices which have undergone these design processes when making purchasing decisions.

The research cited above broadly falls into two types: empirical studies of ethical concerns arising in response to particular smart devices in particular situations, or expository work exploring how long-standing ethical principles might relate to smart homes in general. Any insights gained therefore tend to be either highly specific (e.g. smart meter agents raise concerns about autonomy, as in~\cite{costanza2014doing}), or else highly generic (e.g. smart homes in general threaten principles of accountability~\cite{berman2017social}, or discrimination~\cite{wachter2018normative}). Both specificity and generality have their place; specificity has high ecological validity within the context in question, whereas broad ethical principles are useful for informing policy or design initiatives which have to be adaptable to multiple contexts that cannot be specified in advance. However, they are less useful if we want to understand how ethical concerns might be generated from different kinds of smart devices in different contexts.

Taking an intermediate approach allows for an exploration of the human values supported and hindered by different conceptualisations of smartness (see~\cite{friedman2013value}). Normative ethical principles vary by society and culture, and the extent to which they are embedded in devices is determined by designers (modulated by law and regulation). Previous studies have examined user perceptions of specific devices that might be considered smart (e.g. biosensors~\cite{merrill2019sensing}, conversational agents~\cite{lee2019what}, and  home automation~\cite{nilsson2019breaching}). Meanwhile, design tools such as KnowCards\footnote{\url{github.com/betteriot/betteriot-knowcards}} and Envisioning Cards\footnote{\url{www.envisioningcards.com}} help designers to make abstract concerns more concrete. To this end, we examine how end-users conceptualise `smartness'  beyond specific devices and contexts, and how particular formulations of smartness relate to the particular ethical concerns that arise through device usage.

%For instance, if you reduce a smart meter agent's ability to make energy-saving decisions, would that alleviate user concerns about a loss of autonomy? If so, does a similar phenomenon apply to different kinds of devices, across contexts, or not?

% rewrite the bridge sentence - what smartness means beyond individual devices and contexts, and how ethical concerns operate between devices and contexts

% VSD HERE
% abstract ethical principles mean different things to different cultures
% % modulated by their values
% want to show how abstract ethical principles are embodied in smart devices
% Talk about values not VSD - around the abstract notion of values (rather than VSD as a process)

%"However, in the workdescribed here, we use a broader meaning of the term wherein a value refers to what aperson or group of people consider important in life.1In this sense, people find manythings of value, both lofty and mundane: their children, friendship, morning tea,education,art,awalkinthewoods,nicemanners,goodscience,awiseleader,cleanair.

%\cite{friedman2013value} Value sensitive design and Information Systems

%"Value Sensitive Design distinguishes between usability and human valueswithethicalimport.Usabilityreferstocharacteristicsofasystemthatmakeitworkinafunctional sense, including that it is easy to use, easy to learn, consistent, and recoverseasily from errors (Adler and Winograd, 1992; Nielsen, 1993; Norman, 1988).However, not all highly usable systems support ethical values."

\section{Methodology}
In order to investigate the different ways in which users perceived devices to be `smart' we designed a set of surveys targeted at device owners, followed by interviews with a mixture of smart device users to explore the links between perceptions of smartness and ethical concerns they had concerning smart home devices. To inform our survey and interview design we examined related papers, purposefully casting a broad net with respect to characterising ethical concerns according to the definition of ethics used by Jensen et al.~\cite{jensen2018designing}.

All parts of the study were approved by the institution's ethics review board, and all participants were compensated at or exceeding the UK Living Wage. Materials from the study, including interview transcripts, are available at \url{https://osf.io/c3aes}.

\subsection{Survey: Unpacking Smartness}
Even whilst talking about natural intelligence, the idea of ``smart'' can have a variety of meanings and connotations, making it a prime example of a \emph{suitcase word}~\cite{minsky2006emotion}. In the context of connected and algorithmically augmented devices, however, the adjective has been applied to refer to a much larger set of meanings. In order to start unpacking the different perceptions of smartness in the context of smart home devices, we conducted an online survey to capture user understandings about what makes `smart' devices different to non-smart devices. Based on prior literature we chose six different types of smart home devices covering a wide range of functionalities, locations, and usage contexts in the home: light bulbs, speakers (with voice assistants), security cameras, TVs, kitchen appliances, and thermostats. 

Prior to the main survey we conducted two pilot surveys that directly focused on smartness in smart home devices. Understandably, when presented with more abstract questions about whether they trusted their smart devices or considered them intelligent, participants experienced difficulty disentangling perceptions of smartness (e.g. smartness as intelligence vs ideological or historical constructions). These responses reflect the fact that cultural representations of smart technology are influenced by a mixture of positivist marketing, science fiction, and lived experiences, rather than discrete narratives. To this end, the final survey questions approached smartness from a number of different angles. For example:

\begin{itemize}
    \item Q7 and Q8 explored smartness more as a marketing concept
    \item Q23 considered smartness more as an inherent property of the device
    \item Q13 and Q21 attempted to probe how conceptions of smart devices withstood participants' practical experiences with them
\end{itemize}

 We also included the 5 perceived intelligence questions from the `Godspeed' scale~\cite{bartneck2008measuring}, designed to measure perceived intelligence in a human-robot interaction context, in an attempt to draw out other ways in which users might see devices as being smart. A summary of the survey questions is given in Table \ref{tab:survey}.

\begin{table}
    \caption{Sample of questions used to generate the six surveys}
    \label{tab:survey}
    \centering
    \begin{tabularx}{\columnwidth}{l|X}
    \toprule
        Q7 & Why did you choose your smart [device] over other options? Was it being `smart' a factor? \\
        Q8 & If you were describing your smart [device] to a friend, how would you complete the following sentence: "My smart [device] is like a regular [device], except..." \\
        Q13 & If someone you knew was interested in purchasing a [device], would you try and convince them to get a smart [device] over a regular equivalent? What arguments would you use for or against?\\
        Q21 & Recall a time when your smart [device] didn't do what you expected. Did this change your view of your smart [device]? \\  
        Q23 & How do you think your smart [device] could be smarter in 5 years? 15 years? \\
    \bottomrule
    \end{tabularx}
\end{table}

We ran online surveys on the Prolific Academic platform with 20 participants for each device type. Participants owned the device in question, were 18 or over, and were UK residents. Three researchers independently used thematic analysis to generate initial codes for the 120 responses, which were then discussed and combined to produce eight distinct types of smart functionality. These were used to generate the interview protocol.

\subsection{Interviews: Exploring Ethical Concerns Associated with Dimensions of Smartness}
\label{interview:method}
\begin{figure}
    \centering
    \includegraphics[width=1.0\columnwidth]{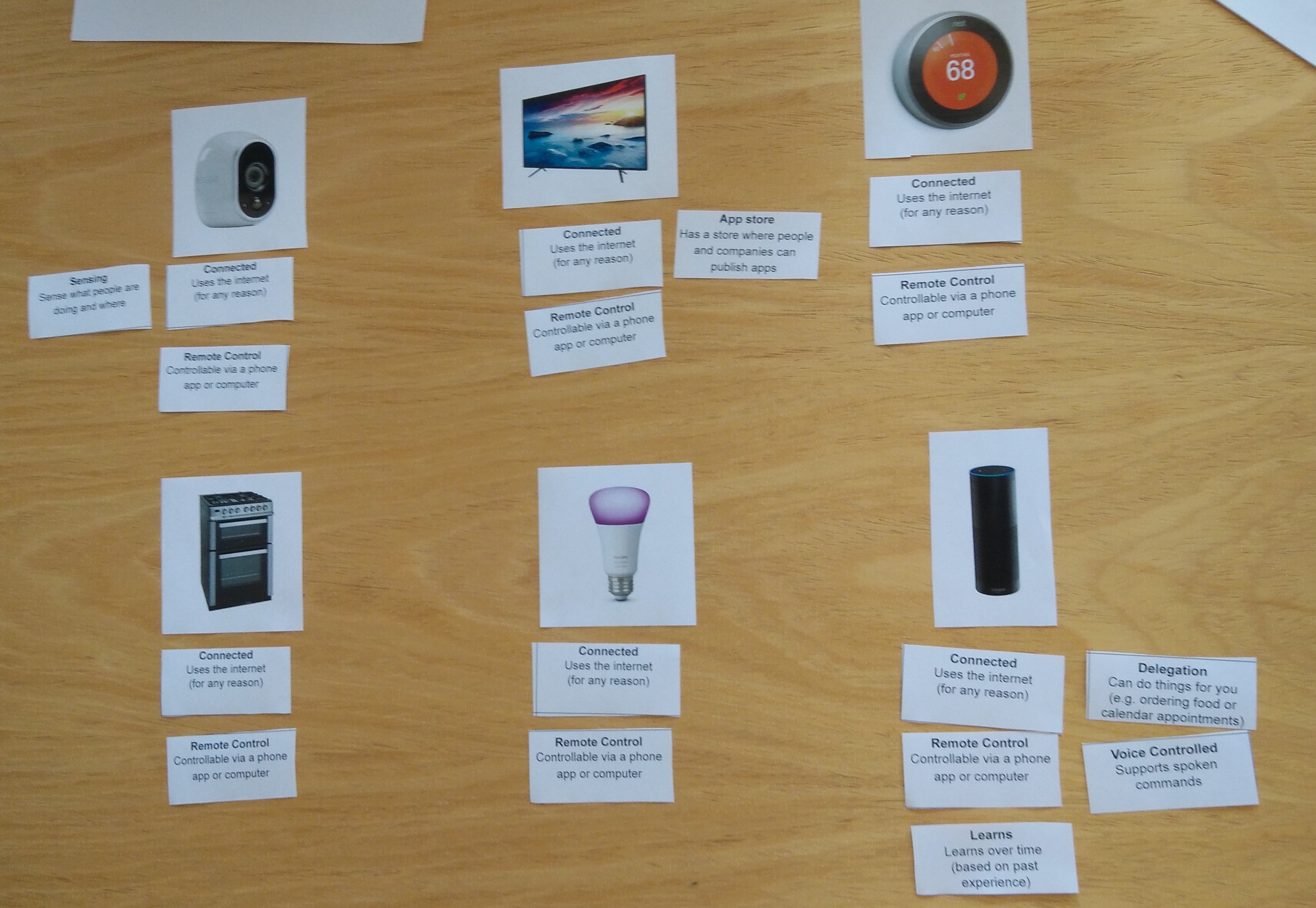}
    \caption{Device and functionality cards after being arranged by [P05].}
    \label{fig:interview}
\end{figure}

The goal of the interviews was to explore how the different dimensions of smartness uncovered in the survey might relate to user's ethical concerns about smart devices. A simplistic approach to examining how perceptions of smartness relate to ethical concerns might be to present these types of smartness individually, and to ask people to respond with ethical concerns for each. However, since our aim was to examine how types of functionality operated to create ethical concerns across contexts, we wanted participants to be able to use the functionalities as a form of mental scaffolding to reason about devices. Inspired by the use of design cards in HCI to stimulate and structure thinking during the design process, we used the types of smartness from the survey analysis to create \textit{functionality cards} that named each type of smartness and gave a brief description (see Figure~\ref{fig:interview}). An additional set of cards depicted the smart home devices used in the survey\footnote{For the kitchen appliance category we initially chose a smart oven, but switched to a smart fridge after two interviews ([P01] and [P05]) due to the oven producing some confusion amongst participants.}. Rather than beginning with abstract concerns and asking how they might apply to products, as in the design process, we did the reverse---beginning with concrete devices and functionalities in order to draw out associated ethical concerns.

% but focuses on concete devices and functionalities to draw out concerns and throughs rather than starting with concerns and working the other way like the design cards would have done

University mailing lists, the Call for Participants platform, and a pool of existing participants were used to recruit participants who had previous experience with smart devices (either their own or via friends and family). As with the surveys, participants were UK residents aged 18 or over. Interviews took place at the university and lasted 30 to 45 minutes.

At the beginning of the interview, the six device cards were arranged on the table in a random order, and the eight functionality cards were placed face down to the side. In order to establish a baseline for each participant's experience with and understanding of smart devices, we began each interview by asking which smart devices they had used, either currently or at some point in the past. We also asked them to identify the devices and describe what they felt made them smart. Participants were introduced to the functionality cards with a brief explanation and asked to arrange cards next devices they thought had that capability (see Figure~\ref{fig:interview}). Participants were then asked if they would have any concerns about using the device as it was, and whether their concerns would be alleviated by putting it in different locations in the house, or by adding or removing functionality cards.

In order to go beyond participants' immediate personal reactions to these devices, we then reset the functionality cards to a standard configuration and asked them to respond to a set of vignettes depicting smart home contexts. Vignettes are sketches of fictional scenarios, used as a research tool to collect situated data on values, beliefs, and norms, common in sociology and more recently applied to HCI~\cite{coughlan2013tailored, jenkins2010putting}. In an in-depth interview, the respondent is invited to imagine the scenario, drawing on his or her own experience. In utilising fictional details, this method is similar to speculative design approaches, which have been used to study views on the ethical implications of technology use~\cite{lawson2015problematising, lindley2017anticipating, van2016computationally}. Relatedly, Nilsson et al. used breaching experiments with scenario-based design and contra-vision to draw out problems with home automation by creating ``provocative views on the home of the future''~\cite{nilsson2019breaching}. 

Our vignettes were drawn from papers on ethical concerns in smart homes. For each of our vignettes, we asked participants how they thought smart devices might be used by the individual characters in the vignette, which device(s) they imagined being most concerned about and why, and whether adding or removing functionality from those devices would alleviate those concerns (represented by placing or removing functionality cards on the table). The four vignettes were as follows:

\begin{enumerate}
    \item[V1] \emph{Parents and teenagers:} A household in which the parents used the smart devices to monitor what their teenage children were doing, adapted from~\cite{choe2012investigating, cranor2014parents, ur2014intruders}.
    
    \item[V2] \emph{Revenue maximisation:} A home where the manufacturers of each device attempted to generate as many revenue streams as possible, adapted from~\cite{buchanan2016british, parkhill2013transforming, rodden2013home}. 
    
    \item[V3] \emph{Vulnerable people:} The installation of the devices into an environment where vulnerable people were living, such as a care home for older people, or an assisted living facility, adapted from~\cite{demiris2008senior, mclean2011ethical, pradhan2018accessibility, pragnell2000market, springer2018play, wilson2015smart}.
    
    \item[V4] \emph{Rental accommodation:} A rental accommodation where each of the devices had been installed by either another tenant or the landlord, adapted from~\cite{mazurek2010access, oulasvirta2012long, zeng2017end}. 
    
\end{enumerate}

Interviews were recorded and transcribed\footnote{The car mechanic example attributed to P11 in the discussion was given after the interviewer believed the participant had finished talking, and is thus recreated from memory (with consent).}, before being coded according to references to devices and types of smartness from the device and functionality cards. Photos were taken of each participant's arrangement of cards and included for reference in the analysis. An iterative thematic coding process was used to identify common themes relating to dimensions of ethical concerns. Three researchers independently coded disjoint subsets of the transcriptions, before convening to consolidate themes and derive a set of joint codes addressing the varieties of ethical concerns. These were re-applied to the data and are presented in the next section organised by types of smartness. The categories of ethical concerns themselves are elaborated on in the discussion.

\section{Results}
\subsection{Survey: What Does Smartness Consist of?}
The 120 survey respondents had an average age of 34 $(\sigma=11)$; 81 identified as female, and 31 as male. Six declined to provide their age and/or gender identity.

As expected, different conceptions of smartness were often entangled. For some, merely being marketed as smart meant that their device must be better in some sense, whereas for other there was more concrete functionality (described below). Smartness was also presented more neutrally as a trade-off (``made me realise how much more complicated it can be compared to just a normal tv'' [tv16]) or even negatively (``a lot of people have this belief that smart must mean better'' [appliance03]).

The survey results made it clear that for certain devices, such as the TV and the speaker, there was an obvious capability or feature that users associated with that device being ``smart''. Looking across the results from different devices also led us to other ways that respondents consistently described smart functionality (such as being able to control the device via a smartphone). Analysis of the survey results yielded eight distinct features, ranging from more concrete to more abstract, that were considered to contribute to device smartness\footnote{It is interesting to note that some of these categories partially correspond to attributes used in European Parliament discussions about smart robots~\cite{urquhart2019responsible}, and practitioner definitions of smart home devices~\cite{kim2019definitions}.} outlined in Table~\ref{tab:smartness}.

\begin{table}
    \centering
    \caption{Types of smart functionality identified from the survey results}
    \label{tab:smartness}
    \begin{tabularx}{\columnwidth}{r|X}
        \toprule
        Functionality & Description \\ \hline
        Apps & Can install apps from an app store \\
        Voice Control & Responds to spoken commands \\
        Connected & Connects to the internet \\
        Remote Control & Via an app or computer \\
        Sensing & Knows what's going on \\
        Responsive & Responds to sensing or other stimuli \\
        Learns & Behaviour is based on past interactions \\
        Delegation & Executes tasks on the user's behalf \\
        \bottomrule
    \end{tabularx}
\end{table}

\textbf{Apps} were the most common smart functionality given for the TV, and the second most commonly mentioned for the smart speaker (after voice control). \textbf{Voice control} was the most commonly cited smart element of respondent's speakers, often mentioned in terms of what users could do with voice control, rather than as a feature in and of itself.

\textbf{Internet connectedness} was rarely mentioned explicitly during the descriptive parts of the survey, but in response to Q14, 83\% of respondents (99) said they believed that their device was connected to the internet, with a further 6\% (7) stating that they were unsure. \textbf{Remote control} was a common theme, particularly amongst the more functional devices such as the thermostat, light bulbs, and kitchen appliances. 

\textbf{Sensing} was commonly mentioned when describing the functionality of the thermostat and camera in Q8: \textit{``It can sense when you are returning and turn itself on'}' [thermostat10]. \textbf{Responding} to stimuli came across from responses about cameras , but was also mentioned frequently in the context of smart speakers (``\textit{It responds well and is able to complete many different tasks''} [speaker12]).

\textbf{Delegation} was a common theme with smart speakers and thermostats, with users mentioning being able to create reminders and purchase products through the device, as well as thermostats being better at controlling the heating than they were. \textbf{Learning} was a popular response to Q23, with suggestions about learning preferred TV programs, learning about people in the house, and even becoming more human-like.

\subsection{Interviews: Connections Between Types of Smartness and Ethical Concerns}
\label{interview:results}
A total of 15 participants were interviewed, ranging from those who had only used others' smart home devices (P15) to those who had all 6 example devices and more (P09). The gender identities and ages of the participants are shown in Table \ref{tab:age-demo}.

\begin{table}
    \centering
    \caption{Demographics of interview participants}
    \label{tab:age-demo}
    \begin{tabular}{c|c|c|c|c|c}
    \toprule
     P\# & Gender Id. & Age & P\# & Gender Id. & Age\\ \hline
     P01 & F & 18-24 & P09 & M & 25-34 \\
     P02 & M & 35-44 &  P10 & M & 25-34  \\
     P03 & M & 25-44 &  P11 & M & 45-54  \\
     P04 & M & 18-24 & P12 & F & 35-44 \\
     P05 & F & 35-44 & P13 & M & 25-34 \\
     P06 & F & 35-44 & P14 & F & 25-34 \\
     P07 & M & 55-64 & P15 & M & 18-24 \\
     P08 & M & 25-34 &&&\\
     \bottomrule
    \end{tabular}
\end{table}

\subsubsection{Delegation}

Among the kinds of tasks and activities for which delegation was seen as most welcome---that is, for which participants expressed the fewest reservations-- were those that could be described as ``monitoring activities''. Delegation in this case was seen as an opportunity to `keep track' of things so that people wouldn't have to. For kitchen appliances, such activities were often safety-related, such as smart ovens detecting or preventing fires. These activities were also discussed for other devices specifically designed for safety or security, such as cameras, including those which facilitated the remote monitoring of elderly relatives. In this way, delegation was seen to enable people to continue living independently who might otherwise need more full-blown paternalistic help:

\begin{quote}
    I've got a friend whose mum has got early onset dementia and she's got [...] sensors in the door. So one day, her mum went to bed and her phone pinged at 11 o'clock saying did you know your mum's left her door open, which is brilliant because she can get her next door neighbour to go and shut it. [P02]
    
    The more automated things are then the more it could be useful for people who don't have a capacity to understand time. [P12]
\end{quote}

%\emph{``the more automated things are then the more it could be useful for people who don't have a capacity to understand time.''} [P12].

Delegating monitoring activities was seen as a means of  enabling people to live more efficiently---by simply making it possible for those who didn't have the time or effort to pay attention to their energy usage:

\begin{quote}
     [A smart thermostat] could be very useful because I know some people they just don't care---they just set 20$\degree$ every day, they waste so much money and energy without even realising. [P05]
\end{quote}

For other contexts, however, participants expressed reservations toward delegation. For some this was down to a puritanical notion that delegation could encourage laziness: \textit{``if everything gets done for you [...] you don't really make any effort. I think I like some effort''} [P05]. More commonly, participants expressed hesitation regarding delegated tasks that had potential value-laden consequences in the real world.  In such cases, there was a concern that devices might make bad choices resulting in loss: \textit{``it could be a surprise if you find loads of rotting food sitting at your doorstep''} [P07]. Communicating or messaging others was also seen to be complex and fraught with risks: \textit{``it's a bit weird that something has so much power as calling someone or sending a message to someone--''} [P05]. However, when the parameters for these these actions were well-specifiable and the mechanisms were perceived as reliable, people were less concerned:

\begin{quote}
    I would find it very useful if tasks that are pretty simple to understand can be [delegated], because then I know these things are on autopilot, almost like a direct debit, so I know it's being dealt with [...] in an OK way. [P11]
\end{quote}

\subsubsection{Learning}
On the one hand, participants were happy, even expectant, that devices learn in certain contexts. These usually related to: \emph{direct learning}---comprising learning by example (such as how to heat a house based on daily adjustments); via explicit instruction (such as mood lighting); or in cases of reminders and recommendation (e.g. \emph{``you'd want your TV to learn what you like to watch''} [P02]). 

But participants expressed being less comfortable when the learning related to information that they had not intentionally given the device, or was about things that were distinctly personal, such as a person's thoughts, activities, or preferences: \emph{``I have reservations about technology learning how we think and how we act''}. [P02]

Some felt that learning preferences was akin to passing judgement (\textit{``I wouldn't want my fridge to judge what I was eating''} [P13]), while some participants made a distinction between learning from the individual's own data in isolation as as opposed to with data from others: 

\begin{quote}
     The huge power of smart devices is that they don't just learn from me, but they learn from millions of people---that's a bit freaky. [P07]
\end{quote}

Participants expressed further reservations for devices learning things they initially did not expect the device was capable of learning, and discovering things learnt caused their perceptions about it to change from positive to creepy or invasive. In many cases, this was because participants saw this kind of extended learning going too far: \emph{``it's like it's got a mind of its own, that weirds me out''}~[P01].

As with delegation, the ability for devices to learn was often contrasted against ideals of control, but also of transparency, with participant's lack of understanding about \textit{how} and \textit{what} devices were learning leading them to cease feeling in control of them: \emph{``You might not want it to be learning. You might want to be in more control of it''}~[P04].

In the context of delegated responsibilities,  learning was also seen as problematic because it could impinge on a person's autonomy--the ability to change and control one's shopping for instance. It was also seen that getting the system to anticipate learn such changes would in some sense cross the line:

\begin{quote}
     Just because you bought it last weekend doesn't mean you want to buy it this week, you might change [what you want to eat]. Either it's going to order things that are not right, or it's going to know far more about me than you would like it to. [P07]
\end{quote}

\subsubsection{Connectedness}
Devices being connected and remote controllable was described as being core to their smartness (in the words of P13, \textit{``that's part of what they are''}), and often one of the key features that drove people to purchase and use them. In addition to being necessary for the delivery of new features, participants also saw it as required for updating devices and any apps that ran on them.

Participants clearly understood that being able to connect out would also mean that others might be able to connect in, making connectedness an obvious locus for concerns about privacy, with all participants describing fears about data being stolen or hackers compromising and controlling devices such as cameras. Often, when asked which functionality cards they would remove to alleviate these concerns, participants would start by touching the connected card before pausing as they realised that doing so would remove a key part of what made the device desirable:

\begin{quote}
    So if we remove the internet connection it's fine. But then, that's the smartness right, that's the smartness gone. [P11]
\end{quote}

In this way, connectedness was seen as an all or nothing concept---a question of \emph{how much} data to send to remote parties, rather than \emph{how many} remote parties to send data to---where removing it would cause devices to cease being smart. Thus the more connected a device was seen to be (i.e. the more data participants believed it was sending over the internet), the greater their concerns about it were. But in other cases, participants suggested that devices might be able to communicate only \textit{within} the home, forgoing the ability to operate them remotely in order to alleviate concerns stemming from it being connected to the internet:

\begin{quote}
    The camera, I think that if it wasn't connected to the internet that it would be good [...] because traditionally people had closed circuit televisions---that was the point of it, that it's closed circuit and no one else can get in. [P02]
\end{quote}

\subsubsection{Voice Control}
Participants' concerns over voice control manifested primarily as issues around privacy, in particular perceptions of such devices as always listening and attending to what was spoken. This was perceived as unsettling, not only because eavesdropping was human-like behaviour they were not accustomed to devices having, but simply by creating a constant feeling of \textit{presence}:

\begin{quote}
    Because Alexa is something that you talk to, it feels like it occupies a room [...] it's not the fact of having the microphone, it's the fact of having the entity that speaks back to you. [P13]
\end{quote}

Voice control was also seen as problematic due to a lack of awareness about what exactly \textit{what} such devices were doing or capable of doing. Such `fear of the unknown' resulted in some assuming that the extent of the data collection via voice control outstripped their understanding: \emph{``You don't know what else it's recording ... they get some useful marketing information, much more than you're even aware of really''} [P07].

Voice control was discussed most positively in the \textit{vulnerable users} scenario, in terms of the value it brought to individuals with special needs. For instance P01's autistic cousin who had too many questions for his mother to cope with, whereas \emph{``if he's asking a machine, it will just keep answering his questions.''} [P01]. Similarly, participants reflected that their own concerns about voice assistants might be offset by their utility if they became less able: \emph{``Forty years from now, I might have difficulty doing things and appreciate something voice controlled [...] but right now I don't''} [P06].

\subsubsection{Sensing \& Responsiveness}
Participants often discussed \emph{sensing} and \emph{responsiveness} in conjunction; for the purposes of the results, we therefore report both here. Sensing capabilities were seen as a risk if they might inadvertently compromise the privacy of others in their environments, including their own family members---\emph{``I wouldn't really want to know what my wife was up to''} [P02]---and that doing so could undermine relationships, demonstrating \emph{``a lack of trust between the parents and the children [...] you've gotta have that kind of trust, otherwise they're not going to respect you.''} [P02].

Privacy risks associated with sensing seemed to be strongly related to their degree of fidelity, or at least the extent to which information about people could be unambiguously determined.  On one hand, sensors with inherently limited capabilities such as motion detectors and occupancy sensors, often associated with smart lighting or thermostats,  gave rise to the fewest concerns.  At the other extreme were sensors with sophisticated inference capabilities, such as smart security cameras with the potential of facial recognition capacities; P07 argued that while it would be acceptable for the smart camera sense the presence of a person, \emph{``if it can sense \emph{who} it is, that would be a bit freaky''} [P07].

\subsubsection{Remote Control}
The convenience of being able to control smart devices remotely was seen as a key benefit of their smartness, especially for home thermostats, lighting and security cameras. But it was also seen as an opportunity for mischief:

\begin{quote}
    Sometimes on holidays, my boyfriend stays home and I connect to the camera. I say `Hello!' and I scare him. [P05]
\end{quote}

Others described using remote control to subvert traditional means of monitoring by family members (particularly in the context of V1), for instance P01 imagined remotely switching off her home lights to deceive her parents into thinking she had already arrived home whilst still out at night.

But this was often accompanied by the realisation that if \textit{they} could control the device, then perhaps others could too: \emph{``with remote control it's not always a given that you're the person who's in control''} [P02]. In addition to harmless mischief, remote control might tempt landlords or overbearing family to interfere in users' lives against their wishes (V4).

\subsubsection{Apps}
App stores were often seen as enabling third party developers to unlock the promised potential behind smart devices, since \emph{``they have many many more ideas about what can be done with a device than the company that makes the device''} [P11].

While this was seen as essential---\emph{``ecosystems like that must exist''} [P13]---the introduction of third parties via app stores was identified as a potential source of concern, by opening up the surveillance capabilities of devices to actors beyond the manufacturer. Some participants worried that this additional functionality might itself enable other types of privacy violations, for instance if a \emph{``rogue app that you could install [on Amazon Echo] to listen and take the microphone and store what's being said.''} [P11]. For these reasons, some suggested removing the app store to ameliorate privacy concerns: \emph{``the app store, if you get rid of that, then big companies won't be able to track all of your habits''} [P04]. 

Moreover, in response to the \emph{revenue maximisation} scenario, participants described a conflict of interest on the part of manufacturers, who would be incentivised to compromise the user experience in an attempt to make money, \emph{``encouraging people to buy more things on the apps store''} [P12].

\section{Discussion}
The previous section identified ethical concerns associated with key forms of `smartness' across different devices through vignettes that grounded their use in particular contexts (summarised in Table~\ref{tab:concerns}). It showed how the ethical concerns of participants around smartness were often centred around the control of knowledge, either for themselves or their devices, and how all behaviours became increasingly uncomfortable as behaviours drifted towards those characterised as human. Another key dimension to smartness was that being `smart' represented devices (and by extension their users) being a part of something bigger, for better or worse. Here we delve into discussion of the ethical concerns themselves, examining the extent to which they persist, vary, and operate among the constituent forms of `smartness' and contexts, and how they relate to previous conceptualisations from the literature.

\begin{table}
    \centering
     \caption{Connection between types of smartness and ethical concerns}
     \small
    \label{tab:concerns}
    \begin{tabular}{l|l}
        \toprule
        Smartness & Ethical Concern  \\ \hline
        Delegation & Autonomy, Transparency, Uncanny Behaviour \\
        Learning & Privacy, Transparency, Uncanny Behaviour \\
        Connectedness & Privacy \\
        Voice Control & Privacy, Autonomy, Uncanny Behaviour \\
        Remote Control & Privacy, Autonomy, Social Order \\
        Sensing \& Responding & Privacy, Autonomy, Transparency, Social Order \\
        Apps & Privacy \\
        \bottomrule
    \end{tabular}
\end{table}

\subsection{Privacy}
\label{sec:privacy}

Privacy was the most prevalent ethical concern, appearing across all types of smartness except delegation. This is rather unsurprising; previous research into smart devices in a variety of contexts has almost invariably brought up privacy (of the 35 papers used to design the vignettes, all mentioned privacy in some form). However, the manner in which privacy arose, the kind(s) of privacy referred to, and the meaning(s) attached to them subtly differed between the different types of smartness.

While \emph{connectedness} had clear privacy implications associated with the possibility of unwanted data exposure, \emph{voice control} raised privacy issues at a more phenomenological level; it was not necessarily just the fact that Alexa could facilitate privacy-violating data flows, but that having a device listening out for spoken commands induced the feeling of being watched by an ``invisible person'' in the room (P14). Through distinctions like this, we can see how different dimensions of smartness may impinge upon privacy at different levels, from \emph{facts} about whether or not certain data is flowing, to \emph{phenomenology} (how it feels), to \emph{preferences} or \emph{norms}~\cite{ohara2016seven}. 

In articulating how \emph{app stores} and \emph{remote control} might threaten privacy, references were made to the ways in which social contexts may dictate the appropriateness of certain information flows and incursions into private spaces \cite{nissenbaum2004privacy}. P05 felt using the remote control on her home web cam to surprise her boyfriend while he was home alone was acceptable within the contextual norms of their relationship, but agreed it would be a violation for a landlord to do the same to a tenant. In a different way, app stores open up the possibility of data sharing relationships with unknown third party actors beyond the manufacturer, as P04 cautioned; as with smartphone app ecosystems, these third parties may come to be seen as social actors potentially violating contextual integrity through opaque leakage of personal information \cite{van2018x, van2017better}. 

\subsection{Autonomy \& Control}
\label{sec:autonomy}

As outlined in the results, delegation, voice control, and sensing/responding gave rise to discussions pertaining to \emph{autonomy}, understood as the ability to make one's own choices without external influence, or `self-governance'~\cite{colburn2010autonomy}. For instance, P07's feeling that delegating tasks to Alexa might restrict his freedom to choose what to eat for dinner or watch on TV, or P05's puritanical concern that too much delegation would eliminate meaningful effort. However, the effects of devices on autonomy were not always understood to be undermining. In some circumstances, as P12 remarked, the ability to delegate to a device or rely on its capacity to sense and respond, could \emph{support} autonomy by enabling people to continue living independently.

%In some circumstances, the ability to delegate to a device, or rely on its capacity to sense and respond, could \emph{support} autonomy by enabling people to live independently who might otherwise need more full-blown paternalistic help: \emph{``the more automated things are then the more it could be useful for people who don't have a capacity to understand time.''} [P12].

%TODO: swap summary of P12's remark

This apparent contradiction---that features like delegation can both undermine and support autonomy---might be explained by reference to subtly different notions of autonomy they appeal to. P07's objection to having his meal choices decided for him reflects autonomy as `negative liberty', or freedom from interference, while P12's example that delegation could support cognitively impaired people to live independently might reflect `positive liberty', understood as the ability to take control of one's own life~\cite{berlin2017two}.

Participants also suffered from a lack of options when exercising control over \textit{how} they used smart home devices. While the main ways that participants talked about resolving tensions around device functionality appeared to mirror exaggerated smartphone tropes of (dis)integration~\cite{harmon2013stories}, for the devices considered in the study these categories really were total, reflecting the all or nothing aspect of device functionality; participants (often implicitly) saw their options as being to `give in', integrate the device, and accept all of the device's features (desirable \emph{and} undesirable), or to dis-integrate, discard the device, and benefit from none of them.

\subsection{Transparency and Accountability}

Similarly to~\cite{nilsson2019breaching}, the issues around transparency and accountability that participants described in the interviews ranged from simple problems of visibility (e.g. what a device was sensing or recording), to more complex questions about why a device might take a particular action (e.g. as a result of learning, or suspected conflicts of interest between the user and the manufacturer).

Voice control and connectivity offered clear examples of the former, with participants frequently expressing confusion over when speakers were recording, or what information they stored about interactions. This echoed prior findings around devices poorly conveying the limits of data collection~\cite{malkin2018can, mcreynolds2017toys}, and the problems of poor mental models about smart devices~\cite{zeng2017end}. A more detailed analysis showed that users often had good mental models about what devices were \emph{capable} of capturing but little knowledge of what they were \emph{actually} capturing (e.g. The Echo \emph{could} record everything, but is it?), and it was from this that transparency concerns arose, rather than a lack of understanding about capability.

Of the more complex questions, while participants did not expect to understand the learning process, the repeated contrasting of learning with control did suggest that the lack of affordances about learning prevented them from building good mental models, giving rise to the feeling of no longer being in control. For P02, who questioned the appearance of adverts after a conversation in the vicinity of a voice assistant, a simple concern about transparency was elevated to a more complex concern about the lengths manufacturers were willing to go to in order to turn a profit. In some cases, it appeared that users had attempted to deal with these transparency problems by being much more sceptical about devices and manufacturers. Examples included P07's suggestion that companies are collecting `far more than you know' and P04's belief that companies would automatically be tracking users, a type of dejected acceptance previously seen in this context~\cite{shklovski2014leakiness}.

\subsection{Conflicts of Interest}

Some types of smartness were connected to potential commercial conflicts of interest that devalued devices' perceived usefulness. Some of these featured quite familiar and prosaic forms of influence, such as adverts on their fridge LCD display, or a standby screen on the TV: \emph{``It would be [...] akin to putting up a corporate poster in your kitchen.''} [P02]. More complex conflicts of interest arose in relation to voice assistants. The fact that they made it easier for users to purchase from one vendor than others was seen by some as going directly against their will, and reduced the usefulness of the device. This is a typical example of what economists term the agency problem~\cite{grossman1992analysis}, a common theme in the literature around smart meters and smart energy grids~\cite{costanza2014doing, goulden2014smart, rodden2013home}. Along these lines, P11 made an analogy to how customers of his car maintenance business would normally go to their preferred mechanic rather than other options that were cheaper or otherwise more convenient, and other participants also saw the potential for becoming trapped in an ecosystem.

There was also an alternative interpretation, which framed this concern in the context of \textit{competition}. Participants suggested that offering reduced options when delegating was acceptable, so long as the provider of a device did not abuse their dominant position in the market: \textit{``A lot depends on if it's the same company that sells you the device, is also the same company that's providing services behind the device''} [P07]. This was mirrored in a more general concern that once users had purchased one device, manufacturers would use their foothold in the home to sell users other devices. Examples included voice assistants for other rooms, or additional thermostats that would only work with the system the user had originally purchased. In these cases participants worried their ability to purchase whichever device they liked in the future would be curtailed by devices being inoperable with each other.

\subsection{Social Order}

As in many previous studies of multi-user device interaction in smart homes (e.g.~\cite{mennicken2016s, porcheron2018voice, zeng2017end}), a prominent set of concerns pertained to devices disrupting or otherwise failing to respect the natural social order(s) of the home. Much as in Zeng et al.'s study~\cite{zeng2017end}, one major class of these concerns dealt with power and control---both within/among inhabitants, as well as relating to actors external to such spaces. In particular, there was a discussion of the potential hazards of the privileges that \emph{owners as administrators} of smart devices (including landlords) might have over inhabitant-users. This was particularly the case where devices conferred asymmetric power or information access and remote control to owners, including sensing/listening capabilities.

From the perspective of the owner with such privileges, respecting the `politics of control' meant devising ways to negotiate or self-regulate the use of devices to prevent accidental breaches of the trust, as in~\cite{porcheron2018voice}. In doing so, participants discussed having to take deliberate action to \emph{avoid} spying on, or otherwise violating the expectations that family members or those subject to surveillance had about them. In some cases, such efforts were made difficult or impossible by devices not adequately supporting appropriate access restrictions, such as smart TVs failing to differentiate among multiple owners, or smart speakers like Alexa keeping unified audit logs of all users' interactions. 

A special class were caring relationships, discussed through our vignettes focused on parents and their children (V1), as well as with elderly relatives (V3). In such relationships, concerns centred around achieving a balance between devices enhancing the safety of the cared for, and risks of undermining their autonomy and privacy in the process. In the parents and teenagers scenario, it was widely agreed among participants that potential harms arising from the use of such devices included them being used in ways that were overly controlling of their children, as P02 described. 

%\emph{``a lack of trust between the parents and the children [...] you've gotta have that kind of trust, otherwise they're not going to respect you.''}[P02]. 

This was in contrast to responses to the elderly relatives vignette, in which participants were most willing to use the various kinds of smartness to support relatives in re-gaining self-sufficiency. Here, potential ethical concerns regarding relatives' autonomy were seen to be outweighed by the advantages these devices afforded. These two vignettes also highlighted the differences in perspective between our participants, and how trying to adopt a one-size-fits-all approach to smartness-related ethical concerns might lead to the marginalisation of certain voices. Whereas the other vignettes saw more focused themes emerge, with participants drawing on shared cultural backgrounds and public discourse, social order concerns were far more likely to be related to highly personal experiences, creating a diverse patchwork of justifications informing what was deemed acceptable.

\subsection{The Uncanny Valley of Social Smart Things}

On the one hand, participants referred to some devices as if they were social entities~\cite{nass1993anthropomorphism}, as observed in several prior studies~\cite{mennicken2016s,purington2017alexa}. Examples of this include P04 worrying about being bossy to Alexa, or P14 describing it as an invisible person in the room. But on the other hand, tasks such as learning and delegation were described as unsettling because they were \emph{too human}, making a device appear to have a \emph{mind of its own}. In this way, devices sometimes fell into an uncanny valley, where they were perceived as social enough to trigger social behaviours in users, but still machine-like enough to create dissonance~\cite{mori1970uncanny}. These uncanny valley responses also relate to the purification behaviours described in~\cite{leahu2013categories}, where the blurring of human and machine causes users to alter their internal categories to eliminate overlaps or grey areas.

Because these sentiments were linked to normative values about the roles that machines should or should not have in people's lives, it is possible that these reactions will diminish over time, as P02 chillingly remarked with advertising on the door of the smart fridge: \textit{``maybe it's just something that we need to get used to. Something that society will just grow to accept''}. The way that voice control was described as lazy mirrors the way in which many other technologies that are now integrated with everyday life began as curiosities; social use of the telephone was originally characterised as `frivolous and unnecessary'~\cite{pantzar1997domestication}. 

\section{Limitations and Future Work}
While it did not prevent major issues and themes arising in the interviews, the exploratory nature of our study meant our participants constituted a relatively small set of independent adults. This narrower set of views risks erasing accounts of smartness from marginalised groups, as evidenced by the willingness of participants to justify the use of smart devices for older adults with little consideration for their concerns. Additionally, there are countless devices and use cases beyond those of which our participants had experienced. While we attempted to mitigate this with vignettes that were applicable to a wide span of social situations and lifestyles, we hope to more fully address both of these limitations in future studies.

\section{Conclusion}
Designers of smart devices are increasingly called upon by researchers, advocates, and policymakers to anticipate and mitigate the negative ethical consequences of their innovations~\cite{berman2017social,fritsch2018calling}. While clearly much-needed and well-motivated, such calls present designers with under-specified goals. Investigations of particular devices in specific contexts of use can provide rich insights into particular user's idiosyncratic concerns \emph{in situ}, and abstract discourse may help ground those concerns in light of longstanding principles. But both approaches are either too specific or too generic; neither enables reflection on how different dimensions encapsulated by the `smart' moniker relate to ethical concerns across devices and contexts.

The analysis presented here is intended to begin exploring the design space of `smartness'. The formulations of smartness given by our participants help cut through the nebulous use of `smart' both in the marketing of these devices and in academic research; while understandings of smartness are unlikely to have neat boundaries (especially as the concept shifts across cultures and time), by outlining some of these rough edges we aim to help designers better anticipate how other devices and contexts might influence ethical concerns about their own. In this way, unpacking smartness allows for its subsequent \textit{repacking} in order to help ``manage the potential `attack surface' of the digital on everyday life''~\cite{crabtree2017repacking}, that is, supporting designers in designing for the values of their users.

\section{Acknowledgements}
This work was funded by EPSRC grant N02334X/1.

\balance{}

\bibliographystyle{SIGCHI-Reference-Format}
\bibliography{main}

\end{document}